\documentclass[12pt]{article} 
\usepackage{amssymb} \usepackage{amsmath} \usepackage{amsfonts} \usepackage{graphicx} \usepackage{mathrsfs} \usepackage[dvips]{color}

   \newcommand{\Z}{\mathbb{Z}} \newcommand{\N}{\mathbb{N}}

 \newcommand{\bn}{{\mathbf{n}}}  
    
 \newcommand{\bE}{\mathbf{E}}

 \newcommand{\cE}{\mathcal{E}}  \newcommand{\cG}{\mathcal{G}}

 \newcommand{\cV}{\mathcal{V}}  
  \newcommand{\be}{\begin{equation}} \newcommand{\ee}{\end{equation}}
\newcommand{\bea}{\begin{eqnarray}} \newcommand{\eea}{\end{eqnarray}} \newcommand{\nn}{\nonumber} 
  
\newcommand{\ed}{\end{document}}

  \newcommand{\bi}{\begin{itemize}} \newcommand{\ei}{\end{itemize}}
    
  \newcommand{\bce}{\begin{center}} \newcommand{\ece}{\end{center}}

  \newcommand{\p}{\partial} \newcommand{\tE}{{\tilde E}}

\begin{document}

\title{Chiral fermions on 2D curved spacetimes}

\author{Farhang Loran\thanks{E-mail address:loran@cc.iut.ac.ir} \\[6pt] Department of Physics, Isfahan University of Technology, \\ Isfahan
84156-83111, Iran\\[6pt] }

\date{ } \maketitle

\begin{abstract}

 The theory of free Majorana-Weyl spinors is the prototype of conformal field theory in two dimensions in which the gravitational anomaly and the
 Weyl anomaly obstruct extending the flat spacetime results to curved backgrounds. In this paper, we investigate a quantization scheme in which
 the short distance singularity in the two-point function of chiral fermions on a two dimensional curved spacetime is given by the Green's
 function corresponding to the classical field equation.  We compute the singular term in the Green's function explicitly and observe that the
 short distance limit is not well-defined in general. We identify constraints on the geometry which are necessary to resolve this problem. On
 such special backgrounds the theory has locally   $c=\frac{1}{2}$  conformal symmetry.
\vspace{2mm}

\noindent PACS number: 04.62.+v, 11.30.Rd, 11.25.Hf, 11.25.Pm \vspace{2mm}

\noindent Keywords: conformal field theory, curved surface, short distance singularity, gravitational anomaly, Weyl anomaly.

\end{abstract}

\section{Introduction}

 In a two dimensional conformal field theory on a curved background, the conformal anomaly, the Weyl anomaly, the gravitational anomaly
 \cite{Alvaerz-Gaume-Witten} (manifested as the loss of the general covariance or the local  Lorentz symmetry \cite{Bardeen-Zumino,
 Alvarez-Gaume-Ginsparg}) and the holomorphic factorization property \cite{Stora91} are closely interconnected.

  In a free field theory on a flat background the  canonical fields are chiral, the Hilbert-space is essentially a Verma module over the
  corresponding
  Virasoro algebra, and the partition functions is holomorphically factorized \cite{Ginsparg}.  In a weak gravitational field one can `expand
  around the flat space' and compute the correlation functions perturbatively \cite{Alvaerz-Gaume-Witten,GSW,Big-Book}. In this way, one can show
  that the anomalies are basically given by the central charge $c$ of the corresponding conformal field theory on a flat background.  One can
  also add local counterterms to shift one anomaly to another one, and use the unobstructed symmetry in order to extend definite aspects of the
  quantum field theory on the flat spacetime to the quantum field theory on a curved background \cite{Stora90, Karakhanian}.

 The original example of a two dimensional field theory with the gravitational anomaly is the
 theory of spin $\frac{1}{2}$ Weyl fermions on a curved surface with Minkowski signature \cite{Alvaerz-Gaume-Witten}.
 In \cite{Leutwyler85, Leutwyler86}, it is shown that in this model, the anomalies can be expressed in terms of
 the short distance singularities of the fermion propagator. Chiral fermions has been extensively used to model
 chiral matter coupling to the two dimensional quantum gravity. See for example \cite{OZ-npb, Ketov} and references
 therein.

 In this paper, we investigate the consistency of a quantum field theory of free Majorana-Weyl spinors on an arbitrary curved background.  From the flat spacetime limit one already knows
 that the leading term in the two-point function scales like $[z(x)-z(x')]^{-1}$ as $x\to x'$ where the {\em coordinate} $z(x)$ labels locally the
 light-like curves. Motivated by the path-integral approach, we assume that the short distance singularity of the two-point function is  given by the short distance singularity of the
 Green's function corresponding to the `original'  classical field equation.\footnote{Focusing on the local properties of quantum fields has been a leading principle in the
 reformulation of quantum field theory on curved spacetimes, {\em cf.} \cite{Axiomatic-QFT}.} By `original' we mean that none of the classical symmetries are used
 to simplify the theory before quantization. Thus we compute the singular part of the Green's function on an arbitrary curved background
 explicitly and examine the limit $x\to x'$.
  Since in two dimensions the {\em two}-vector $(x-x')$ has two components, the limit $(x-x')\to0$ corresponds to two successive limits in two
  different directions. In general, these limits do not necessarily exist or commute with each other at the singularity. Resolving this ambiguity
  imposes a constraint on the geometry which can be expressed in terms of a simple condition on the tetrad ${e_\mu}^a$: either
 \be
 \p_\mu \left[\left(\det e\right) {e^\mu}_\pm\right]=0,
 \label{Main-result-1}
 \ee
 where $\pm$ denote the light-like directions, or there should be a globally well-defined solution  $V_\rho=V_{\rm gl}$ to the differential
 equation
 \be
 {V_\rho}^{-1}{e^\mu}_\pm\p_\mu V_\rho=\left(\det e\right)^{-1}\p_\mu \left[\left(\det e\right) {e^\mu}_\pm\right].
  \label{Main-result-2}
 \ee
 We  call geometries that satisfy the conditions \eqref{Main-result-2} and \eqref{Main-result-1} special of the first kind and special of the
 second kind respectively. On such backgrounds one recovers the anticipated singularity. Furthermore one can consistently define locally
 conformal field theories whose central charge equal $\frac{1}{2}$.

 {Chiral fermions on a Ricci flat surface are free of anomalies  \cite{Alvaerz-Gaume-Witten, Witten}. In appendix \ref{App-BG2nd} we show that a special manifold which is of the second kind in both light-like directions is Ricci flat.  This result can be justified by noting that taking the limit $x\to x'$ of the correlation function amounts to moving a spinor field along a certain path. Thus the limit is well-defined, i.e.,  path-independent only if the curvature is zero.\footnote{We are grateful to D. Karakhanyan for pointing out this simple observation.} From this point of view, short distance singularity of the  scalar propagator on a curved background is expected to be generally well-defined. This will be confirmed rigorously in section \ref{section-scalar}.}

 {The special manifolds of the first kind are in general curved, but, as we show in section \ref{special-models} the existence of $V_{\rm gl}$ ensures that there exist a globally well-defined map to a chiral theory on a Ricci flat surface, which is not obstructed by anomalies.  Some instructive examples of such geometries are given in section \ref{subsect-examples}.}

 The organization of the paper is as follows. In section \ref{Notation}, we collect our notation and conventions. Section \ref{section-scalar} is
 an advertisement for the quantization scheme, where we reproduce the standard results in the massless scalar field theory. In section
 \ref{section-fermion} we use the quantization scheme to study the Majorana-Weyl fermions on a curved background. The conditions given in
 Eq.\eqref{Main-result-1} and Eq.\eqref{Main-result-2} are
 derived in section \ref{Fermi-Statistics}.  Our results are summarized in section \ref{Summary}. Some technical details are relegated to
 appendices.

\section{Notation and conventions}\label{Notation}
 $g_{\mu\nu}$ denotes the metric on the curved surface and $\eta^{ab}$ stands for the Minkowski metric. Throughout the paper, $\eta^{ab}$ is in
 the light-cone gauge, i.e. $\eta=\boldsymbol{\sigma}_1$. $\boldsymbol{\sigma}_i$  are the  Pauli matrices. The zwei-bein ${e_\mu}^a$ and its
 inverse ${e^\mu}_a=\eta_{ab}\,g^{\mu\nu}{e_\nu}^b$ describe the non-coordinate basis
 \be
 g^{\mu\nu}=\eta^{ab}\,{e^\mu}_a\, {{e^\nu}}_b={e^\mu}_+{{e^\nu}}_-+{e^\mu}_-{{e^\nu}}_+,
 \ee
 where $g^{\mu\nu}$ is the inverse of $g_{\mu\nu}$. $e:=\sqrt{-\det g}$ and ${E^\mu}_a:=e\,{ e^\mu}_a$. $\delta_{\rm D}$ stands for the Dirac
 delta function. ${\boldsymbol{\delta}}$ denotes the Kronecker delta function and ${\boldsymbol{\varepsilon}}:=i\boldsymbol{\sigma}_2$. These
 symbols appear either by the Latin indices or the Greek indices.  We will use ${\boldsymbol{\delta}}_{\mu\nu}$ to raise and lower the indices of
 ${\boldsymbol{\varepsilon}}$ and ${E^\mu}_a$ i.e. $E_{\mu a}:={E^\mu}_a$. It should be noted that $E_{\mu a}\neq e e_{\mu a}= g_{\mu\nu}
 {E^\nu}_a$. In section \ref{section-fermion} one encounters ${\tE^\mu}_{\ \pm}:={{\boldsymbol{\varepsilon}}^{\mu}}_\nu {E^\nu}_\pm$. In
 components, $({\tE^1}_{\ \pm},{\tE^2}_{\ \pm})=({E^2}_\pm,-{E^1}_\pm)$. Denoting ${E^\mu}_\pm$ collectively by ${E^\mu}$  one verifies that $
 {{E_\mu}}{\tE^\mu}=0$, and
 \be
 \bE^2:={E_\mu} {E^\mu}={{\tE}_\mu} {{\tE}^\mu}.
 \ee
  Dirac matrices are denoted by $\gamma^a$. $\gamma^0=\boldsymbol{\sigma}_1$ and $\gamma^1=-i\boldsymbol{\sigma}_2.$

\section{Scalar field theory}\label{section-scalar}
 The massless scalar field theory is given by the action
 \be
 S=\frac{1}{2}\int d^2x \sqrt{-\det g}g^{\mu\nu}\p_\mu\phi\p_\nu\phi.
 \label{scalar-action-original}
 \ee
 The field equation $\partial_\mu\left(\sqrt{-\det g}g^{\mu\nu}\p_\nu\phi\right)=0$  in terms of the tetrad $e_\mu ^{\ a}$ reads
 \be
 \partial_\mu\left(e\,{{e^\mu}_a}\,{{e^\nu}}_b\,\eta^{ab}\,\partial_\nu\,\phi\right)=0.
 \label{scalar-FE-original}
 \ee
  We seek local chirality in the sense of, locally,
  \be
  \phi(x)=\phi_-\left[z_-(x)\right]+\phi_+\left[z_+(x)\right],
  \label{ver1-phi-pm}
  \ee
   such that $\p_\pm\phi_\mp=0$, where
  \begin{align}
  \label{p+-}
  &\p_\pm:=f_\pm {e^\mu}_\pm\partial_\mu.
  \end{align}
 The functions $z_\pm(x)$ are  locally determined by the equations
  \begin{align}
  \label{z+-}
 &  \p_az_b={\boldsymbol{\delta}}_{ab},&a,b=\pm.
  \end{align}
 The functions $f_\pm$ in Eq.\eqref{p+-} are  (partially) determined by the condition $[\partial_+,\partial_-]=0$.  Since the filed equation
 \eqref{scalar-FE-original} reads
 \be
 \partial_\mu\left(e\,{{e^\mu}_a}\,{{e^\nu}}_b\,\eta^{ab}\,\partial_\nu\,\phi\right)=\p_\mu\left(\frac{{E^\mu}_+}{f_-}\right)\p_-\phi+
 \p_\mu\left(\frac{{E^\mu}_-}{f_+}\right)\p_+\phi+\frac{2e}{f_+f_-}\p_+\p_-\phi=0,
 \label{scalar-FE}
 \ee
 the necessary condition for  the local chirality is
 \be
 \p_\mu\left(\frac{{E^\mu}_\pm}{f_\mp}\right)=0,
 \label{necessary-condition}
 \ee
 which is readily fixed by the condition $[\partial_+,\partial_-]=0$, see Appendix \ref{App-A}. By construction, $\partial_\pm$ are generators of
 light-like curves. For example, for $\delta x^\mu=\epsilon\p_+ x^\mu$,
 \be
 g_{\mu\nu}\delta x^\mu\delta x^\nu=\epsilon^2 e^2f_+^2 g_{\mu\nu}e^\alpha_+e^\beta_+\p_\alpha x^\mu\p_\beta x^\nu=\epsilon^2
 e^2f_+^2\eta_{++}=0.
 \ee
 Thus $\phi_+(z_+)$ and $\phi_-(z_-)$ correspond to the locally  left-mover and the locally right-mover components of the scalar field
 respectively.

 In the context of the classical field theory, it is traditional to describe this result in terms of the celebrated property of the two
 dimensional geometries. In two dimensions, metric has three independent components, which, upon coordinate transformation, can be given
 (locally) in terms of a single function, known as the conformal factor. In the present case, definition \eqref{p+-} implies that
 \be
 \p_\pm x^\mu=f_\pm {e^\mu}_\pm.
 \label{diff-scalar}
 \ee
 Therefore, metric in the $z_\pm$ coordinates equals $f_+f_-\eta_{\mu\nu}$. So the conformal factor equals $f_+f_-$. Since the canonical scalar
 field $\phi$ is (classically) invariant under diffeomorphism, $\phi(x^\mu)\to\Phi(z_\pm)=\phi\left[x^\mu(z_\pm)\right]$ one easily verifies
 that
 \be
 S=\int d^2z \p_+\phi\p_-\phi.
 \label{scalar-flat-action}
 \ee

 \subsection{The quantization scheme}
 Since the anomalies could possibly ruin the classical symmetries, one should be cautious about using Eq.\eqref{scalar-flat-action} to quantize the classical theory \eqref{scalar-action-original}.  The first step toward quantization is the identification of the short distance singularity in the operator
 product expansion  of the canonical fields $\phi_\pm$ defined in Eq.\eqref{ver1-phi-pm}.  The relevant term in the operator product
 expansion is determined by the short distance singularity of the propagator. In the path integral approach, the propagator is given by the  Green's function corresponding to the field equation \eqref{scalar-FE-original}
 \be
  \partial_\mu\left[e\,{{e^\mu}}_a\,{{e^\nu}}_b\,\eta^{ab}\,\partial_\nu\, G_{\rm sc}(x,x')\right]=\delta_{\rm D}^2 (x-x'),
 \ee
 in which $\delta_{\rm D}^2 (x-x')$ denotes the Dirac delta function.  Using  the first equality in  \eqref{scalar-FE}, Eq.\eqref{necessary-condition} and the identity
 \be
 \delta_{\rm D}^2(x-x')=\frac{e}{f_+f_-} \delta_{\rm D}\left[z_+(x)-z_+(x')\right]\delta_{\rm D}\left[z_-(x)-z_-(x')\right],
 \ee
 which can be deduced from  Eq.\eqref{diff-scalar}, one verifies that
 \be
 \p_+\p_-G_{\rm sc}\left[z_\pm(x),z_\pm(x')\right]=\frac{1}{2}\delta_{\rm D}\left[z_+(x)-z_+(x')\right]\delta_{\rm
 D}\left[z_-(x)-z_-(x')\right].
 \ee
 This equation is enough for determining  the short distance singularity of the Green's function without identifying the boundary conditions. Consequently,
 \be
 \lim_{x\to x'}G_{\rm sc}(x,x')= \frac{i}{4\pi}\lim_{x\to x'}\left\{\ln\Big[z_+(x)-z_+(x')\Big]+\ln\Big[z_-(x)-z_-(x')\Big]\right\},
 \label{scalar-Green}
 \ee
 which results in the well-known operator product expansions
 \begin{align}
 \label{scalar-correlation}
& \lim_{x\to x'} \phi_\pm\left[z_\pm(x)\right]\phi_\pm\left[z_\pm(x')\right]=\lim_{x\to x'} \left\{-\ln\Big[z_\pm(x)-z_\pm(x')\Big]\right\},\\
 \label{scalar-left-right-correlation}
 & \lim_{x\to x'} \phi_\pm\left[z_\pm(x)\right]\phi_\mp\left[z_\mp(x')\right]=0,
 \end{align}
 that could be directly obtained from Eq.\eqref{scalar-flat-action}.
 \subsection{Local conformal field theory}
 Eq.\eqref{scalar-left-right-correlation} indicates that, locally, there is zero correlation between the left-movers and the right-movers. Let
 $(z,\p,\phi)$ denote $(z_+,\p_+,\phi_+)$ and $(z_-, \p_-,\phi_-)$ in the left-mover and the right-mover sectors respectively. Define,
 \be
 T(z):=-\frac{1}{2}\lim_{z\to w}\left[\p\phi(z)\p\phi(w)+\frac{1}{(z-w)^2}\right].
 \ee
 Using Eq.\eqref{scalar-correlation}  one easily verifies  that
 \begin{align}
 &\p\phi(z)\p\phi(w)=-\frac{1}{(z-w)^2}+\cdots,\\
& T(z)\p\phi(w)=\frac{\p\phi(w)}{(z-w)^2}+\frac{\p^2\phi(w)}{z-w}+\cdots,\\ &T(z)T(w)=\frac{1/4}{(z-w)^2}+\frac{2T(w)}{(z-w)^2}+\frac{\p
T(w)}{z-w}+\cdots.
 \end{align}
 These short distance behaviors are characteristics of the $c=1$ conformal field theory with a primary field $\p\phi(z)$ with conformal weight
 $h=1$. The generators of the Virasoro algebra can be defined locally in terms of the Laurent expansion of the $T(z)$,
 \be
 T(z)=\sum_{k\in\Z}z^{-k-2}L_k.
 \label{Laurent}
 \ee
 Once more, this construction is in agreement with the traditional approach to the conformal field theory on a flat geometry, starting from the
 action \eqref{scalar-flat-action}. In that case, $T_\pm$ correspond the (anti)holomorphic components of the energy-momentum tensor
 \be
 T^{\mu\nu}=\frac{2}{\sqrt{-\det g}}\frac{\delta S}{\delta g_{\mu\nu}}.
 \ee

 \section{Majorana-Weyl fermions}\label{section-fermion}
 The action principle for the massless Dirac fermion $\psi$ in two dimensions is given by
 \be
 S=\frac{1}{2}\int d^2x\, e{e^\mu}_a\left(\bar\psi\gamma^a\overleftrightarrow{\p_\mu}\psi\right).
 \label{Dirac-action}
 \ee
 In two dimensions, the spin connection drops out of the Lagrangian due to the classical Fermi statistics. The field equation is
 \be
\gamma^a\left[ {E^\mu}_a\p_\mu+\frac{1}{2}\p_\mu {E^\mu}_a\right]\psi=0.
 \ee
Assuming that
  \begin{align}
 \label{psi-components}
 & \psi=\left(\begin{array}{c}u_-\\u_+\end{array}\right),
 \end{align}
 the field equation reads
 \be
 \left[ {E^\mu}_\pm\p_\mu+\frac{1}{2}\p_\mu {E^\mu}_\pm\right]u_\mp=0.
 \ee
  The components $u_+$ and $u_-$ are known as the left-handed and the right-handed Weyl spinors respectively.
  \be
  u_\pm=\frac{1\mp\gamma^0\gamma^1}{2}\psi.
  \ee
  Decomposing $u_\pm=\theta_\pm^{(r)}+i\theta_\pm^{(i)}$ in terms of their real and imaginary components, one verifies that a single massless
  Dirac field $\psi$ is equivalent to four decoupled copies of the Majorana-Weyl spinors $\theta^{(A)}_\pm$, $A=r,i$.
 \be
 \left[ {E^\mu}\p_\mu+\frac{1}{2}\p_\mu {E^\mu}\right]\theta=0,
 \label{theta-FE}
  \ee
  in which, ${E^\mu}$ denotes ${E^\mu}_+$ (${E^\mu}_-$) in the right-handed (left-handed) sector. On a flat geometry, a left-handed fermion is
  left-mover and a right-handed fermion is right-mover. The field equation \eqref{theta-FE} stems from the action \eqref{Dirac-action} which,
  after inserting Eq.\eqref{psi-components} in it, reduces to a sum of four terms similar to
 \be
 S=\frac{1}{2}\int d^2x {E^\mu} \theta{\p_\mu}\theta.
 \label{theta-action}
 \ee
 This theory is invariant under local `Lorentz' transformation
 \begin{align}
  \label{V-transforation-E}
 & {E^\mu}\to {E_V}^\mu=V^{-1}{E^\mu},\\
 \label{V-transforation-theta}
  &\theta\to \theta_V=V^{1/2}\theta,
 \end{align}
 which can be verified by inserting $E_V^\mu$ and $\theta_V$ into either Eq.\eqref{theta-action} or Eq.\eqref{theta-FE} and noting that
 $\theta(x)$ is a Grassmann field i.e. $\theta(x)^2=0$. This identity can be interpreted as the classical Fermi statistics. From the
 transformation rule Eq.\eqref{V-transforation-theta} one also infers that $\theta$ is a spin-$\frac{1}{2}$ field. In the following, we
 generalize this theory a little bit. The new theory is  defined  by  the field equation
 \begin{align}
 &\left[ {E^\mu}\p_\mu+\frac{1}{n}\p_\mu {E^\mu}\right]\theta=0,&n\in\N,
 \label{theta-FE-n}
  \end{align}
  in which $\theta$ is a  Grassmann field  obeying the Lorentz transformation rule
  \be
  \theta\to \theta_V={U_V}^{\frac{1}{n}}\theta.
  \label{U-transforation-theta}
  \ee
 $U_V$ is related to $V$ by the condition
  \be
  {U_V}^{-1}{E^\mu}\p_\mu U_V=V^{-1}{E^\mu}\p_\mu V.
  \label{UV}\ee
 Since the field equation is linear in the background field ${E^\mu}$, one  gains nothing from generalizing ${E^\mu}$ to a  complex field.
 \subsection{Local chirality in the classical theory}
 Let $V_\rho$ denote a special (classical) Lorentz transformation identified by the following equation,
 \be
 {V_\rho}^{-1}{E^\mu}\p_\mu V_\rho=\p_\mu {E^\mu}.
 \label{VR}
 \ee
 Using Eq.\eqref{V-transforation-E}, Eq.\eqref{theta-FE-n} and Eq.\eqref{U-transforation-theta} one easily verifies that
 \begin{align}
 & \p_\mu {E^\mu}_{V_\rho}=0,\\
 \label{local-chirality-classical}
 & {E^\mu}\p_\mu \theta_{V_\rho}=0.
  \end{align}
  For $n=2$ this is the celebrated classical Poincare-Weyl symmetry of massless Dirac fermions in two dimensions.
  Eq.\eqref{local-chirality-classical} conveys the classical local chirality of the field $\theta_{V_\rho}$. To show it, we define operators
  $\p$ and $\bar\p$ as follows.
 \begin{align}
 \label{pbar}
 &\bar\p:=\bar f {E^\mu}\p_\mu,\\
 \label{p}
 &\p:=f{\tE^\mu}\p_\mu.
 \end{align}
 We also need  two real-valued functions $z(x)$ and $\bar z(x)$ locally determined by the equations
 \begin{align}
 \label{ver1-z-barz}
 & \p z=\bar\p \bar z=1,&\bar\p z=\p\bar z=0.
  \end{align}
 It is easy to verify that $\bar\p$ is the generator of a light-like curve given by the equation $dz=0$. Furthermore, in the left-handed and the
 right-handed sectors $\bar\p$ is proportional to, respectively, $\p_-$ and $\p_+$ defined in section \ref{section-scalar}. The functions $f$ and
 $\bar f$ in Eq.\eqref{p} and Eq.\eqref{pbar}  are partially determined by the condition $[\p,\bar \p]=0$ which gives
 \begin{align}
  \label{f}
 &f^{-1}{E^\mu}\p_\mu f=-{\tilde E}_\mu C^\mu,\\
  \label{fbar}
  &{\bar f}^{-1}{\tE^\mu}\p_\mu \bar f={E_\mu} C^\mu,
 \end{align}
 in which,
 \be
 C^\mu:=\frac{1}{\bE^2}\left[{E^\nu}\p_\nu {\tE^\mu}-{\tE^\nu}\p_\nu {E^\mu}\right].
 \ee
 Equations \eqref{f} and \eqref{fbar} can be used to show that under  the Lorentz transformation  \eqref{V-transforation-E}
   \begin{align}
   \label{V-trans-f}
 &  f\to f_V=VW_V\left[z(x)\right]f,\\
    \label{V-trans-fbar}
 &  \bar f\to f_V=V\bar W_V\left[\bar z(x)\right]\bar f,
  \end{align}
  where, $W_V$ and $\bar W_V$ are arbitrary functions of $z(x)$ and $\bar z(x)$ respectively. See Appendix \ref{App-V-transform-C} for details.
  Eq.\eqref{V-transforation-E}, definitions \eqref{p} and \eqref{pbar}, and transformation rules \eqref{V-trans-f} and \eqref{V-trans-fbar} give
  \begin{align}
   \label{V-trans-p}
 &  \p\to \p_V=W_V\left[z(x)\right]\p,\\
    \label{V-trans-pbar}
 &  \bar \p\to \bar\p_V=\bar W_V\left[\bar z(x)\right]\bar \p.
  \end{align}
  Let
  \be
  \rho:=\bar f\p_\mu {E^\mu},
  \ee
  in terms of which, the field equation \eqref{theta-FE-n} becomes,
  \be
  \left(\bar\p+\frac{\rho}{n}\right)\theta=0.
 \label{theta-FE-new}
   \ee
  Eqs.\eqref{V-transforation-E} and \eqref{V-trans-fbar} indicate that, under the Lorentz transformation $V$,
  \be
  \rho\to\rho_V=\bar W_V\left(\rho-V^{-1}\bar\p V\right).
  \label{V-transforation-rho}
  \ee

  \subsection{Special models}\label{special-models}
  Assume that $\rho\neq0$. Let
  \be
  T^{(\alpha)}:=F^{(\alpha)}\theta\p\theta,
  \label{T-alpha}
  \ee
  where, $\theta$ is on-shell, i.e. the field equation \eqref{theta-FE-n} is assumed. The function $F^{(\alpha)}$ is partially determined by the
  condition
  \be
  \left(\bar\p+\alpha\rho\right)  T^{(\alpha)}=0.
  \label{T-alpha-property}
  \ee
  We also require that under the Lorentz transformation \eqref{V-transforation-E} and \eqref{U-transforation-theta}
  \be
  T^{(\alpha)}\to  T^{(\alpha)}_V=W^{(\alpha)}_V(z)  T^{(\alpha)},
  \label{T-alpha-V-trans}
  \ee
  for some function $W^{(\alpha)}_V$ of $z(x)$. In appendix \ref{App-chiralT} we show that this condition requires,
  \be
  \alpha V^{-1}\bar\p V=0.
  \label{alpha-V}
  \ee
 Let
 \be
 \cV_{ch}=\{V|V^{-1}{E^\mu}\p_\mu V=0\},
 \label{VH}
 \ee
 denote the (sub)group of `chiral' Lorentz transformations. We call a theory defined by the field equation \eqref{theta-FE-new} a `chiral model'
 if the classical symmetry group is chiral i.e. both $V$ and (as a result of Eq.\eqref{UV}) $U_V$ belong to $\cV_{ch}$. Otherwise, the theory is called
 non-chiral. The condition \eqref{alpha-V} indicates that in the non-chiral models, $\alpha=0$.

 In Appendix \ref{App-Special-Lorentz} we find a solution to Eq.\eqref{VR}  or equivalently  ${V_\rho}^{-1}\bar \p V_\rho=\rho$.
 If $V_\rho$ is well-defined all-over the spacetime, we denote it by $V_{\rm gl}$.
  $V_{\rm gl}\not\in \cV_{ch}$ and corresponds to a global Lorentz transformation.
 In the chiral models the classical symmetry group consists entirely of the chiral Lorentz transformations
 $V\in\cV_{ch}$. Consequently, $V_{\rm gl}$ is a not a member of the classical symmetry group of chiral models. So there is no obstruction  (even
 after quantization) in using $V_{\rm gl}$ to replace the canonical field $\theta$ and the background ${E^\mu}$ by new ones
 \begin{align}
 \label{replacement}
 &\theta\stackrel{V_{\rm gl}}{\to}\vartheta:={V_{\rm gl}}^{\frac{1}{n}}\theta,& {E^\mu}\stackrel{V_{\rm gl}}{\to}{\cE^\mu}:={V_{\rm gl}}^{-1}{E^\mu},
 \end{align}
 satisfying the identity $\p_\mu{\cE^\mu}=0$ and the field equation
 \begin{align}
 {\cE^\mu}\p_\mu\vartheta=0.
 \label{vartheta-FE}
 \end{align}
  We shall call such a chiral model a `special chiral model'.   In such models one can define, without loss of generality,
  $T(z):=\vartheta\p\vartheta$.  To put in  a nutshell, if $\rho\neq0$ and a globally well-defined Lorentz transformation $V_{\rm gl}$ exists,
  then the geometry supports  chiral Majorana-Weyl fermions $\vartheta$. In the following we call such geometries `special backgrounds of the
  first kind'.

 It should be noted that in the non-chiral models even if $V_{\rm gl}$ exists, a replacement (Lorentz transformation) similar to
 Eq.\eqref{replacement} could be obstructed especially by the Weyl anomaly.

 We shall also call backgrounds with $\rho=0$, `special backgrounds of the second kind'.
 In such backgrounds, $\theta$ is a locally chiral field, $\bar\p\theta=0$ and, inspired by the definition
 \eqref{T-alpha}, $T=\theta\p\theta$. 
 One realizes that  special backgrounds of the second kind are a subclass of special backgrounds of the first
 kind. In particular, Eq.\eqref{V-transforation-rho} indicates that the Lorentz
 symmetry of the chiral spinors on the special backgrounds of the
 second kind is given by $\cV_{ ch}$.

 \subsection{The correlation function}\label{Fermi-Statistics}
 Motivated by the path-integral approach, we assume that the short distance singularity in the two-point function
 $\left\langle\theta(x)\theta(x')\right\rangle$ is given by the short distance singularity of the Green's function corresponding to the field equation \eqref{theta-FE-n}. We give the details in section \ref{App-Fermion-Green}. For now, since we are focusing on the short distance singularities, it is reasonable to use the local coordinates $z$ and $\bar z$ defined in Eq.\eqref{ver1-z-barz} and study the field equation in its simpler form given by Eq.\eqref{theta-FE-new}. In these coordinates, the short distance singularity of the Green's function is encoded in $G_{\rm sp}(x,x')$, defined by the equation
 \be
 \left(\bar\p+\frac{\rho}{n}\right)G_{\rm sp}(x,x')=-2\pi i \delta_{\rm D}\left[z(x)-z(x')\right]\delta_{\rm D}\left[\bar z(x)-\bar z(x')\right].
 \ee
 The Dirac delta function on the right hand side of this equation is defined through the identity
 \be
 \int_{\mathcal{B}} \delta_{\rm D}\left[z(x)-z(x')\right]\delta_{\rm D}\left[\bar z(x)-\bar z(x')\right] dzd\bar z=1,
 \ee
 where $\mathcal{B}$ is a neighborhood of  $x'$. By using the identity $\bar\p(z-z')^{-1}=2\pi i\delta_{\rm D}(z-z')\delta_{\rm D}(\bar z-\bar {z'})$ one can show that
 \be
 G_{\rm sp}(x,x')=-\frac{\exp\left[-\frac{1}{n}\left(R(x)-R(x')\right)\right]}{z(x)-z(x')},
 \label{Green-fermion-bare}
 \ee
 in which $\bar \p R=\rho$. Therefore, 
 \be
  \lim_{x\to x'}
 \left\langle\theta(x)\theta(x')\right\rangle=\lim_{x\to x'}G_{\rm sp}(x,x').
 \label{Fermi-statistics-2-point}
 \ee  
 The existence of the limit $x\to x'$ requires that $ \rho=0$. This claim can be verified by inserting the Taylor series
 \be
 R(x)-R(x')=\left[z(x)-z(x')\right]\p R(x')+\left[\bar z(x)-\bar z(x')\right]\bar \p R(x')+\cdots,
 \ee
 in Eq.\eqref{Green-fermion-bare} which gives
  \be
  \lim_{x\to x'}\left[G_{\rm sp}(x,x')+\frac{1}{z(x)-z(x')}\right]=\lim_{x\to x'}\left[\frac{\bar \p R(x')}{n}\left(\frac{\bar z(x)-\bar z(x')}{
  z(x)- z(x')}\right)+\cdots\right].
  \ee
  The right hand side of this equality is well-defined only if $\bar \p R(x')=0$. We discuss this technicality further in section
  \ref{App-Fermion-Green}.

  All in all, the quantization scheme \eqref{Fermi-statistics-2-point} works only for the special backgrounds
 of the second kind where, by definition, $\rho$ is identically zero. This is a tough restriction on imaginable
 backgrounds. Following our discussion in section \ref{special-models}, one can nullify this condition substantially
 (but not completely) by considering the special chiral models. Such  models, as a matter of fact, are restricted to the special backgrounds of
 the first kind.

 If the background is not special of the first kind, one can still define the fields  $\vartheta$ locally, and at each coordinate neighborhood
 quantize the theory according to the quantization scheme described here, separately. Nevertheless, the (gravitational and Weyl) anomalies
 obstruct fitting together the quantum theories defined on two overlapping patches. It should be noted that although there are no such
 obstructions on special backgrounds of the first kind, there is still no guarantee for the existence of a globally well-defined quantum field
 theory on them.
   \subsubsection{The short distance singularity}\label{App-Fermion-Green}
 In this section we elaborate on Eq.\eqref{Fermi-statistics-2-point} and repeat the above argument explicitly in the $x$-coordinates. Here, the
 short distance singularity of the two-point function is given by the short distance singularity of  the Green's function $\cG_{\rm sp}(x,x')$
 corresponding to the field equation \eqref{theta-FE-n}
 \be
 \left[ {E^\mu}\p_\mu+\frac{1}{n}\p_\mu {E^\mu}\right]\cG_{\rm sp}(x,x')=-2\pi i e(x')\delta_{\rm D}^2(x-x').
 \ee
  Since
 \be
 \delta_{\rm D}\left[z(x)-z(x')\right]\delta_{\rm D}\left[\bar z(x)-\bar z(x')\right]=e\,f\,\bar f\,\bE^2\,\delta_{\rm D}^2(x-x'),
  \ee
  which can be inferred from definitions \eqref{pbar} and \eqref{p},  and using Eq.\eqref{fE2} one verifies that
  \be
  \cG_{\rm sp}(x,x')=-\frac{\left[f(x)\bE^2(x)\right]^{-\frac{1}{n}}\left[f(x')\bE^2(x')\right]^{\frac{1}{n}-1}}{z(x)-z(x')}.
  \label{n2-fake}
  \ee
  Essentially, $\cG_{\rm sp}(x,x')=\left[f(x')\bE^2(x')\right]^{-1}G_{\rm sp}$ {\em cf.} Eq.\eqref{Green-fermion-bare}. $\cG_{\rm sp}(x,x')$ just
  captures  the short distance singularity in the correlation function and is ignorant of the large scale requirements e.g. the boundary
  conditions. The complete correlation function is given by
  \be
  \cG(x,x')=\cG_{\rm sp}(x,x')+\cG_{\rm ns}(x,x'),
  \ee
  where $\cG_{\rm ns}(x,x')$ is a non-singular function. This function is determined by the large scale properties of the Green's function
  $\cG(x,x')$ such as the boundary conditions and the spin structure, and by the Fermi statistics $\cG(x,x')=-\cG(x',x)$.

  Since the short distance singularity in $\cG(x,x')$ is given by the short distance singularity in $\cG_{\rm sp}(x,x')$ i.e.
  \be
  \lim_{x\to x'}\cG(x,x')=\lim_{x\to x'}\cG_{\rm sp}(x,x'),
  \label{lim-lim}
  \ee
  one should carefully examine the existence of the limit $x\to x'$ in Eq.\eqref{n2-fake}.  Since
 \be
 \left(x-x'\right)^\mu=  \frac{\left(x-x'\right)^\nu E_\nu}{\bE^2} {E^\mu}+ \frac{\left(x-x'\right)^\nu \tE_\nu}{\bE^2} {\tE^\mu},
 \ee
 the limit $x\to x'$ encodes two successive limits $\sigma\to 0$ and $\tilde\sigma\to 0$ where
 \begin{align}
 & \sigma:=\left(x-x'\right)^\nu E_\nu,&\tilde\sigma:=\left(x-x'\right)^\nu \tE_\nu.
 \end{align}
 In the presence of a singularity, these two limits do not necessarily commute with each other and the limit $x\to x'$ is not well-defined. In
 particular  $\lim_{\tilde\sigma\to0}\lim_{\sigma\to0}\frac{\sigma}{\tilde\sigma}=0$
 but $ \lim_{\sigma\to0}\lim_{\tilde\sigma\to0}\frac{\sigma}{\tilde\sigma}$ is ambiguous.\footnote{$\lim_{\sigma\to {\
 ^\pm0}}\lim_{\tilde\sigma\to0}\frac{\sigma}{\tilde\sigma}=\pm\frac{1}{0}$!} The limit $x\to x'$ in Eq.\eqref{n2-fake} is subject to the same
 ambiguity: Using Eq.\eqref{d1} and Eq.\eqref{d2}, one verifies that
  \be
  z(x)-z(x')=\frac{\tilde\sigma}{f(x')\bE^2(x')}+\cdots.
  \ee
  Therefore the short distance singularity in $\cG_{\rm sp}$ is given by,
  \be
  \lim_{x\to x'}\cG_{\rm sp}(x,x')=
  -\lim_{x\to x'}\frac{\left[f(x)\bE^2(x)\right]^{-\frac{1}{n}}\left[f(x')\bE^2(x')\right]^{\frac{1}{n}}}{\tilde\sigma}.
  \label{ver2-lim}
  \ee
   The numerator equals
  \be
  1-\frac{1}{n}\left(x-x'\right)^\mu\frac{\p_\mu\left(f\bE^2\right)}{f\bE^2}+\cdots=1-\frac{1}{n{\bE^2}}\left(\sigma
  E^\mu+{\tilde\sigma}\tE^\mu\right)\frac{\p_\mu\left(f\bE^2\right)}{f\bE^2}+\cdots
 \label{numerator}
   \ee
 Using Eq.\eqref{numerator} in Eq.\eqref{ver2-lim} one obtains
 \be
  \lim_{x\to x'}\left[\cG_{\rm sp}(x,x')+\frac{1}{\tilde\sigma}\right]=\lim_{x\to x'}\left\{\left[\frac{1}{n{\bE^2}}
  E^\mu\frac{\p_\mu\left(f\bE^2\right)}{f\bE^2}\right]\frac{\sigma }{\tilde\sigma}+\cdots\right\}.
 \ee
 Consequently the limit $x\to x'$ is well-defined only if
  \be
  {E^\mu}\p_\mu\left(f\bE^2\right)=0,
  \ee
  which, using Eq.\eqref{fE2}, implies that $\rho=0$.

     For special backgrounds of the first kind, one can repeat all these steps after using the replacement \eqref{replacement}.
 \subsection{Local $c=\frac{1}{2}$ Virasoro algebra}
  On special backgrounds one can define the operator $T$ by
 \be
 T(z):=\lim_{z\to w}\frac{1}{2}\left[\Theta(z)\p\Theta(w)+\frac{1}{(z-w)^2}\right],
 \ee
 where $\Theta=\theta$ on the special backgrounds of the second kind and $\Theta=\vartheta$ (defined in Eq.\eqref{replacement}) in the special
 chiral models defined on the special backgrounds of the first kind. One easily verifies that generators $L_k$ defined by the Laurent  expansion
 \eqref{Laurent} satisfy the Virasoro algebra with central charge $c=\frac{1}{2}$. Since $\Theta$ has conformal weight $h=\frac{1}{2}$, one
 realizes that for $n=2$, i.e. for the theory defined by the action principle Eq.\eqref{theta-action},  $T(z)$ generates the corresponding chiral
 Lorentz transformations $\cV_{ch}$, {\em cf.} Eq.\eqref{V-transforation-theta}.
 \subsection{Examples}\label{subsect-examples}
 The first example is a special background. Assume that
 \be
 ds^2=e^{-\omega(x)}\left(dt^2-dx^2\right).
 \label{discuss-1}
 \ee
 where $\omega(x)$ is a smooth function. Let
 \begin{align}
 & {e^\mu}_\pm=2^{-\frac{1}{2}}e^{s_\pm(x)}\left({\boldsymbol{\delta}}^\mu_t\pm{\boldsymbol{\delta}}^\mu_x\right),
 \label{discuss-2}
 \end{align}
where,  $s_\pm(x)$ are smooth functions satisfying the equality $s_-(x)+s_+(x)={\omega(x)}$.
 If e.g. $s_+=\omega$, then $\p_\mu {E^\mu}_+=0$ and the right-handed spinor `experiences' a special background of
 the second kind. In the left-handed sector, $\p_\mu {E^\mu}_-=-2^{-\frac{1}{2}}\partial_x\exp(-\omega)$.
 Since $\omega$ is a smooth function,  $V^{(-)}_{\rm gl}=\exp(-\omega)$ is globally well-defined. Therefore,
 the background in the left-handed sector is special of the first kind.
  Following Eq.\eqref{replacement}, the corresponding `effective' background is given by
 \be
 {\cE^\mu}_\pm=2^{-\frac{1}{2}}\left({\boldsymbol{\delta}}^\mu_t\pm{\boldsymbol{\delta}}^\mu_x\right),
 \ee
 where we have used $V^{(+)}_{\rm gl}=1$ in the definition ${\cE^\mu}_\pm:={V^{(\pm)}_{\rm gl}}^{-1}{E^\mu}_\pm$ .
 The corresponding line element is
 \be
 ds^2=dt^2-dx^2.
 \label{discuss-3}
 \ee
 If $s_\pm\neq \omega$, a similar argument shows that both of the left-handed and the right-handed sectors are
 special chiral models with classical symmetry groups $\cV^{(\pm)}_{ch}=\{V|V^{-1}(\p_t\pm\p_x)V=0\}$ on the flat
 geometry Eq.\eqref{discuss-3}.  All in all, we have realized that the left-handed and the right-handed Weyl
 spinors on the smooth geometry \eqref{discuss-1} classically enjoy the chiral symmetry groups $\cV^{(\pm)}_{ch}$
 and simulate Weyl spinors on the flat background \eqref{discuss-3}.\footnote{There are alternative approaches to Dirac fields on (A)dS$_2$
 spacetimes which can be
 modeled by Eq.\eqref{discuss-1}. See \cite{Epstein2016} and references therein.}

 In the next two examples, we study geometries which are not special but have simply connected subspaces special of the first kind.
 The first one is a two dimensional spacetime given by
 \begin{align}
 & ds^2=d\theta^2-(\sin \theta)^2d\phi^2,&\theta\in[0,\pi],\ &\phi\in[0,2\pi).
 \end{align}
 One easily verifies that ${E^\mu}_\pm=2^{-\frac{1}{2}}\left[(\sin
 \theta){\boldsymbol{\delta}}^\mu_\theta\pm{\boldsymbol{\delta}}^\mu_\phi\right]$, thus, $\p_\mu {E^\mu}_\pm=2^{-\frac{1}{2}}\cos\theta$. The
 special Lorentz transformation is given by $V_\rho=\sin\theta$ which is only well-defined (i.e. it is invertible) for $\theta\in(0,\pi)$. So, if
 one discards the `singular events' at $\theta=0,\,\pi$, and restricts the geometry to the interval $\theta\in(0,\pi)$, one can define a special
 chiral model on a background given by
 \be
 {\cE^\mu}_\pm=\frac{1}{\sin\theta}{E^\mu}_\pm.
 \ee
  The corresponding line element is
 \begin{align}
 & ds^2=dt^2-d\phi^2,&t\in(-\infty,\infty),\ &\phi\in[0,2\pi),
 \end{align}
 where $t=\ln\tan\frac{\theta}{2}$. Therefore the corresponding special chiral model (on both of the left-handed and
 the right handed sectors) becomes equivalent to the ordinary Majorana-Weyl model on the Minkowski cylinder.

 The second example is the Schwarzschild black hole in two dimensions.
 \begin{align}
& ds^2=f(x)\,dt^2-\frac{dx^2}{f(x)},&f(x)=1-\frac{x_0}{x}.
 \end{align}
 $x_0$ denotes the event-horizon.  Let's restrict the geometry to the subspace $x>x_0$. Now $V_{\rm gl}=\sqrt{f(x)}$ is well-defined all-over the
 spacetime. ${{\cE^\mu}_\pm}$ corresponds to the line element
 \begin{align}
 & ds^2=dt^2-dX^2,&X\in(-\infty,\infty),
 \end{align}
 where $X=x+x_0\ln\left(\frac{x}{x_0}-1\right)$.

 Finally, consider the background given by  \cite{KPZ}
 \begin{align}
& {E^\mu}_+\p_\mu=\p_+-h\,\p_-,&{E^\mu}_-\p_\mu=\p_--h\,\p_+.
 \end{align}
 This background is  special of the second kind  in the left-handed  and the right-handed sectors if $\p_+h=0$ and $\p_-h=0$ respectively.

 \section{Summary}\label{Summary}
 We have investigated a quantization scheme in which the short distance singularity in the operator product expansion of the canonical fields is
 determined by the short distance singularity in the Green's function corresponding to the classical equation of motion. We observed that the
 free massless scalar field theory enjoys local $c=1$ conformal symmetry on every smooth two dimensional geometry. For the Majorana-Weyl
 fermions, we realized that this quantization scheme is consistent  only if the two dimensional geometry is a special background (of the second
 kind) in the sense that
 \be
 \p_\mu {E^\mu}=0,
 \ee
  in which
  \be
   {E^\mu}:=\bn^a\left(\det e\right) {e^\mu}_a,
   \ee
  where ${e_\mu}^a$ denotes the tetrad and $\bn$ is a unit vector indicating the light-like direction  corresponding to the handedness of the
  Majorana-Weyl fermion.

 If $\p_\mu {E^\mu}\neq 0$ but the geometry is special of the first kind i.e. it supports a globally  well-defined solution to the equation
 \be
 {V_{\rm gl}}^{-1}{E^\mu}\p_\mu V_{\rm gl}=\p_\mu {E^\mu},
 \ee
  one can consistently define special chiral Majorana-Weyl fermions $\vartheta_n$, $n\in\N$. Such theories are chiral in the sense that
 \begin{align}
 & {E^\mu}\p_\mu\vartheta_n(x)=0,
& V(x)^{-1}{E^\mu}\p_\mu V(x)=0,
 \end{align}
 where $V(x)$ denotes the (chiral) local Lorentz transformation
 \begin{align}
& {E^\mu}\to V(x)^{-1}{E^\mu}, &\vartheta_n\to V(x)^{1/n}\vartheta_n.
 \end{align}
 The classical symmetry group on special backgrounds of the second kind is larger than the chiral group and includes every local Lorentz
 transformation.  Majorana-Weyl fermions on  special  backgrounds retain locally the $c=\frac{1}{2}$ conformal symmetry.

  Our results do not assure the existence of a globally well-defined  quantum field theory of Majorana-Weyl spinors on special backgrounds.
  Investigating the complete set of the necessary and sufficient conditions is beyond the scope of the present work. A promising starting point
  could be identifying a globally well-defined vacuum state via the local conformal symmetries.

 \appendix
 \section{Local chirality: massless scalars}\label{App-A}
 Following the definition \eqref{p+-}, the condition $[\p_+,\p_-]=0$ requires that
 \be
 \frac{\p_\pm f_\mp}{f_\pm f_\mp}=\mp  e^\mp_\nu \left(e_+^\mu\p_\mu {e^\nu}_- - e_-^\mu\p_\mu {e^\nu}_+\right)=e^{-1}e_\pm^\mu\p_\mu e+\p_\mu
 {e^\mu}_\pm=e^{-1}\p_\mu {E^\mu}_\pm,
 \label{app-A1}\ee
 where, to obtain the second equality we have used the identity
 \be
 \p_\mu e=-ee^a_\nu\p_\mu {{e^\nu}}_a.
 \ee
 The last equality in Eq.\eqref{app-A1} gives Eq.\eqref{necessary-condition} immediately.
 \section{Local chirality: Majorana-Weyl spinors}\label{App-V-transform-C}
 Eq.\eqref{V-transforation-E} gives the Lorentz transformation of background field $C^\mu$.
 \be
 C^\mu\to C^\mu-V^{-1}\left(\frac{{\tE^\mu} {E^\nu}-{E^\mu}{\tE^\mu}}{\bE^2}\right)\p_\nu V.
 \ee
 Using Eq.\eqref{f} and Eq.\eqref{pbar}, one verifies that
 \be
 f_V^{-1}\bar\p f_V= f^{-1}\bar\p f+ V^{-1}\bar\p V,
 \ee
 which implies Eq.\eqref{V-trans-f}. Eq.\eqref{V-trans-fbar} can be obtained similarly.
 \section{Chiral $T$-fields}\label{App-chiralT}
 Eq.\eqref{T-alpha-property} gives,
 \be
 \left[\bar\p+\left(\alpha-\frac{2}{n}\right)\rho\right]F^{(\alpha)}=0,
 \label{F-alpha}
 \ee
 where we have used the identity $\theta(x)^2=0$.  On the one hand, Eq.\eqref{F-alpha} together with Eq.\eqref{V-trans-pbar} and
 Eq.\eqref{V-transforation-rho} gives
  \be
 \frac{\bar\p F^{(\alpha)}_V}{ {F^{(\alpha)}_V}}=\frac{\bar \p F^{(\alpha)}}{{F^{(\alpha)}}}+\left(\alpha-\frac{2}{n}\right)V^{-1}\bar \p V.
 \label{app-b-1}
 \ee
 On the other hand the requirement \eqref{T-alpha-V-trans} together with Eqs.\eqref{V-trans-p} and \eqref{U-transforation-theta} gives
 \be
 F^{(\alpha)}\to F^{(\alpha)}_V=W^{(\alpha)}(z){W(z)}^{-1}{U_V}^{-2/n}F^{(\alpha)}.
  \label{app-b-2}
  \ee
  Using Eq.\eqref{app-b-2} in Eq.\eqref{app-b-1}, and using Eq.\eqref{UV} one obtains Eq.\eqref{alpha-V}.

 \section{Special Lorentz transformation}\label{App-Special-Lorentz}
 Definitions \eqref{p} and \eqref{pbar} give
 \bea
 \label{d1}
 \p_1&:=&\left(f\bar f\bE^2\right)^{-1} \left(fE^1\bar \p+\bar f E^2\p\right)\\
  \label{d2}
  \p_2&:=&\left(f\bar f\bE^2\right)^{-1} \left(fE^2\bar \p-\bar f E^1\p\right).
 \eea
 Consequently,
 \bea
 \p_\mu {E^\mu}&=&\left(f\bar f\bE^2\right)^{-1} \left(\frac{1}{2}f\bar\p\bE^2+\bar f{\tE_\mu}\p {E^\mu}\right)\nn\\
 &=&\left(f\bar f\bE^2\right)^{-1}\bar \p\left(f\bE^2\right),
 \eea
 where, we have used Eq.\eqref{f} to obtain the second equality. Let $V_\rho:=f\bE^2$. Obviously
 \be
 {V_\rho}^{-1}\bar\p V_\rho=\rho.
 \label{fE2}
 \ee
  So, if there is a function $f$ such that $V_\rho$ is globally well-defined, the geometry is special of the first kind and  supports  special
  chiral Majorana-Weyl fermions. Eq.\eqref{V-trans-f} and Eq.\eqref{V-transforation-E} show that $f\bE^2$ is invariant under the chiral Lorentz
  transformations only if $W_V(z)=V(z)$.
  \section{Special backgrounds of the second kind}\label{App-BG2nd}
 In this section, we identify  backgrounds which are special of the second kind in both of the left-handed and the right-handed sectors i.e.
 \be
 \p_\mu {E^\mu}_\pm=0.
 \ee
 The most general solution to this equation is
 \be
 {E^\mu}_\pm=\frac{1}{\sqrt 2}{\boldsymbol{\varepsilon}}^{\mu\nu}\p_\nu K_\pm.
 \ee
 for arbitrary functions $K_\pm(x)$. This gives,
 \begin{align}
 & g_{tt}=-\dot K_+\dot K_-,& &g_{tx}=-\frac{K'_+\dot K_-+K'_-\dot K_+}{2},&& g_{xx}=-K'_-K'_+.
 \end{align}
 The corresponding scalar curvature is zero. Therefore, the theory is free of gravitational anomalies and can be consistently quantized.
 The ansatz of static gravitational fields, which are special of the second kind in both sectors, is
 \be
 ds^2=dt^2-F(x)^2dx^2.
 \ee
 A static gravitational field that is special in the left-handed sector is also special in the right-handed sector or vice versa.

\ed
 \subsection{Discussion}\label{Discussion}
 In general relativity (at least in the metric formulation) the canonical fields are the metric components, and the tetrad is an auxiliary field.
 On the contrary, in the quantization scheme that we have described in this article, the background geometry is given by the tetrad ${e_\mu}^a$.
 Therefore one obtains different quantum field theories for different choices of the tetrad. This strange aspect of the quantization scheme is a
 reflection of the Lorentz anomaly \cite{Leutwyler85}. We take it for granted that since fermions are essentially a representation of the local
 Lorentz transformations, they are sensitive to the tetrad instead of the metric.